\documentclass[aps,onecolumn,showpacs,nofootinbib,showkeys]{revtex4-2}
\usepackage[margin=3.0cm]{geometry}
\RequirePackage[T1]{fontenc}

\usepackage{epsfig,graphicx,amsmath,amssymb,bm}
\RequirePackage{mathptmx}      % use Times fonts if available on your TeX system
\usepackage{mathrsfs}
\usepackage{amssymb}
\RequirePackage{color}
\usepackage{changes}
\usepackage{slashed}
\usepackage{multirow}
\usepackage{scalerel}
\usepackage{tikz-feynman}
\usepackage{textcomp}
\usepackage{subcaption}
\usepackage[english]{babel}%используем русский и английский языки с переносами
\usepackage{float}
\RequirePackage{hyperref}
\hypersetup{
    linktocpage,
    colorlinks,
    citecolor=blue,
    filecolor=black,
    linkcolor=blue,
    urlcolor=blue,
}
\usepackage{nccmath}

\begin{document}

\title{Decays $\tau \to \pi \pi \eta\nu_\tau$ and $\tau \to \pi\eta\eta\nu_\tau$ in the extended Nambu--Jona-Lasinio model}


\author{M.K. Volkov$^{1}$}\email{volkov@theor.jinr.ru}
\author{A.A. Pivovarov$^{1}$}\email{pivovarov@theor.jinr.ru}
\author{K. Nurlan$^{1,2,3}$}\email{nurlan@theor.jinr.ru}

\affiliation{$^1$ Bogoliubov Laboratory of Theoretical Physics, JINR, 
                 141980 Dubna, Moscow region, Russia \\
                $^2$ The Institute of Nuclear Physics, Almaty, 050032, Kazakhstan}   

%%%%%%%%%%%% ABSTRACT %%%%%%%%%%%%
%%%%%%%%%%%%%%%%%%%%%%%%%%%%%

\begin{abstract}
In the framework of the extended Nambu--Jona-Lasinio model, the processes $\tau \to \pi \pi \eta(\eta')\nu_\tau$ and $\tau \to \pi\eta\eta(\eta')\nu_\tau$ are considered taking into account mesons in the ground and first radially excited intermediate states. It is shown that in the processes $\tau \to \pi \pi \eta(\eta')\nu_\tau$ the vector channel is dominant, and in the processes $\tau \to \pi\eta\eta(\eta')\nu_\tau$ the main contribution is given by the axial vector channel. The scalar meson $a_0$ plays a dominant role in processes with two $\eta$-mesons in the final state. The significance of the relative phase between the ground and first radially excited states for these processes is shown. 
The obtained results for the $\tau \to \pi \pi \eta\nu_\tau$ process are in satisfactory agreement with the recent experimental data from BaBar and CMD-3, which differ from the averaged values given in the PDG tables.

%\keywords{}

\end{abstract}


\maketitle

\section{Introduction}
The processes related to the nonperturbative region of the theory of strong interactions are the objects of intensive research in modern electron-positron collider centers such as SLAC (BaBar), KEK (Belle), BEPC II (BES III), VEPP (SND, CMD-3) and others. It is important to note the current experiments of BES III, Belle II, and the planned super Charm - Tau factories that add relevance to the study of the processes of strong interactions. The planned $\tau$ decay experiments at the lepton colliders, particularly at the super Charm - Tau factory, provide high statistics and accuracy. The study of processes of meson production in $\tau$ decays is a good laboratory for the analysis of strong interactions effects at low energies.

However, the theoretical description of the processes in this energy region causes difficulties related to the inapplicability of the QCD perturbation theory. The spectral functions and the meson $\tau$ decay widths can be calculated on the basis of phenomenological models taking into account the quark structures of mesons and the chiral symmetry of strong interactions. Among such models one can note the chiral perturbation theory \cite{Gasser:1983yg,Gasser:1984gg} and the Nambu--Jona-Lasinio (NJL) model \cite{Nambu:1961tp,Eguchi:1976iz,Ebert:1982pk,Volkov:1984kq,Volkov:1986zb,Ebert:1985kz,Vogl:1991qt,Klevansky:1992qe,Hatsuda:1994pi,Ebert:1994mf,Volkov:2005kw}, and various extended versions of these models. 

The NJL model is based on the chiral symmetry of strong interactions that is partially broken by the current masses of the $u$, $d$ and $s$ quarks. When considering the $\eta$ mesons, it is important to take into account the mixing of the light $u$ and $d$ quarks with a heavier $s$ quark. Such mixing appears as a result of accounting for the gluon anomaly that is well described in the NJL model by using the 't Hooft interaction \cite{tHooft:1976rip,Volkov:1998ax}. In the description of the decays of the $\tau$ lepton with the production of the $\eta/\eta'$ mesons, the effect of chiral symmetry breaking plays an important role.

One can note the recent success of the NJL model in the description of a series of four particle $\tau$ decays with three pseudoscalar mesons with kaons and $\eta$ mesons in the final states \cite{K:2023kgj,Volkov:2023pmy,Volkov:2023evp}.
 
Currently, the process $\tau \to \pi \pi \eta\nu_\tau$ is actively studied experimentally. In recent experiments for this decay, the results have been obtained that differ from the values given in the PDG: $Br(\tau \to \pi \pi \eta\nu_\tau)_{BaBar} = (1.63 \pm 0.08) \times 10^{-3}$ \cite{BaBar:2018erh}, $Br(\tau \to \pi \pi \eta\nu_\tau)_{CMD-3} = (1.68 \pm 0.17) \times 10^{-3}$ \cite{Gribanov:2019qgw}, and  $Br(\tau \to \pi \pi \eta\nu_\tau)_{PDG} = (1.39 \pm 0.07) \times 10^{-3}$ \cite{ParticleDataGroup:2024cfk}. This gives relevance to the theoretical study of processes of this type. The $\tau \to \pi \pi \eta(\eta')\nu_\tau$ decays are closely related to the electron-positron annihilation processes, since the vector channel is decisive in both cases. Therefore, the widths of the $\tau$ decays can be estimated from the $e^+e^-$ data using the vector current conservation hypothesis (CVC) \cite{Cherepanov:2011zz}. However, for a deeper understanding of the internal structure of the decays, it is definitely of interest to provide direct calculations.

The present work is devoted to the description of the decays $\tau \to \pi \pi \eta(\eta')\nu_\tau$ and $\tau \to \pi\eta\eta(\eta')\nu_\tau$ in the NJL model. For this purpose, both the ground and first radially excited meson states are taken into account as intermediate resonances. Besides, when considering the processes $\tau \to \pi\eta\eta(\eta')\nu_\tau$, the scalar states are taken into account, the contribution of which turns out to be significant, unlike most other processes of this type. An important role is played by taking into account the phase factor, which determines the phase shift between the ground and first radially excited states of mesons. The presence of such a phase factor was established in experimental studies of meson production processes in $e^+e^-$- annihilation \cite{BaBar:2018erh,Gribanov:2019qgw,SND:2014rfi}.

\section{Lagrangian of the NJL model}
The extended NJL model used in the present work was formulated in \cite{Volkov:1996br,Volkov:1996fk}. In this model, quark-meson Lagrangian was obtained as a result of the mixing of the ground and first radially excited meson states:
\begin{eqnarray}
	\label{Lagrangian}
		\Delta L_{int} & = &
		\bar{q} \biggl[ \sum_{j = \pm} \lambda_{j}^{\rho} \left(A_{a_0}{a_0}^{j} + B_{a_0}\hat{a_0}^{j}\right) +
		i \gamma^{5} \sum_{j = \pm} \lambda_{j}^{\pi} \left(A_{\pi}{\pi}^{j} + B_{\pi}\hat{\pi}^{j}\right) +
		+\frac{1}{2} \gamma^{\mu} \sum_{j = \pm} \lambda_{j}^{\rho} \left(A_{\rho}\rho^{j}_{\mu} + B_{\rho}\hat{\rho}^{j}_{\mu} \right) \nonumber \\ 
		&& 
		+\frac{1}{2} \gamma^{\mu}\gamma^{5} \sum_{j = \pm} \lambda_{j}^{\rho} \left(A_{a_1}a_{1\mu}^{j} + B_{a_1}\hat{a}_{1\mu}^{j} \right)
		+ i\gamma^{5} \sum_{M = \eta, \eta'} \lambda_{u} a^{u}_{M} M
		\biggl]q,
\end{eqnarray}
where $q$ is a multiplet of the $u$ and $d$ quarks with the masses $m_u \approx m_d = m = 270$~MeV, $\lambda$ are the linear combinations of the Gell-Mann matrices. The first radially excited states are marked with a hat. The factors $A_M$ and $B_M$ are defined through the mixing angles:
\begin{eqnarray}
\label{verteces1}
	A_{M} = \frac{1}{\sin(2\theta_{M}^{0})}\left[g_{M}\sin(\theta_{M} + \theta_{M}^{0}) +
	g'_{M}f(k_{\perp}^{2})\sin(\theta_{M} - \theta_{M}^{0})\right], \nonumber\\
	B_{M} = \frac{-1}{\sin(2\theta_{M}^{0})}\left[g_{M}\cos(\theta_{M} + \theta_{M}^{0}) +
	g'_{M}f(k_{\perp}^{2})\cos(\theta_{M} - \theta_{M}^{0})\right],
\end{eqnarray}
where $M$ designates the appropriate meson. The values of the mixing angles are given in Table~\ref{tab_mixing}.
\begin{table}[h!]
\caption{The values of the mixing angles of the ground and first radially excited mesons.}
\begin{center}
\begin{tabular}{ccccc}
\hline
   & $\pi$ & $\rho$ & $a_1$ & $a_0$ \\
\hline
$\theta_M$	& $59.48^{\circ}$	&  $81.80^{\circ}$  & $81.80^{\circ}$ & $72.0^{\circ}$  \\
$\theta^0_M$	& $59.12^{\circ}$	& $61.50^{\circ}$  & $61.50^{\circ}$ & $61.50^{\circ}$  \\
\hline
\end{tabular}
\end{center}
\label{tab_mixing}
\end{table}

In the case of the $\eta$ mesons, the four states are mixed: $\eta$, $\eta'$, and their first radially excited states. In this regard, the factors $a_\eta$ take the following form:
\begin{eqnarray}
    a^{u}_{\eta} & = & 0.71 g_{\eta} + 0.11 g'_{\eta} f(k_{\perp}^{2}), \nonumber\\
    a^{u}_{\eta'} & = & -0.32 g_{\eta} - 0.48 g'_{\eta} f(k_{\perp}^{2}).
\end{eqnarray}

The scalar mesons are considered here as quark-antiquark chirally symmetric partners of the pseudoscalar mesons. 

The form factor $f(k_{\perp}^2) = 1 + d k_{\perp}^2$ was included in the model to describe the first radially excited meson states; $d$ is the slope parameter determined from the requirement that the inclusion of excited states does not change the quark condensate.

The coupling constants of the initial non-physical mesons with quarks appear as a result of the renormalization of the Lagrangian:
\begin{eqnarray}
\label{Couplings}
 g_{\pi} = g_{\eta}=\left(\frac{4}{Z_{\pi}}I_{2}\right)^{-1/2}, &\quad&
\, g'_{\pi}=g'_{\eta} =  \left(4 I_{2}^{f^{2}}\right)^{-1/2}, \nonumber\\
g_{\rho} = g_{a_1} =\left(\frac{2}{3}I_{2}\right)^{-1/2}, &\quad&
\, g'_{\rho} = g'_{a_1} =\left(\frac{2}{3}I_{2}^{f^{2}}\right)^{-1/2} \nonumber\\
\, g_{a_0} = \left(4I_{2}\right)^{-1/2}, &\quad& 
g'_{a_0} = \left(4I_{2}^{f^2}\right)^{-1/2},
\end{eqnarray}
where $Z_\pi = \left(1 - 6 \frac{m^2}{M_{a_1}^2}\right)^{-1}$ is the additional renormalization constant appearing while taking into account the $\pi-a_1$ transitions.

The integrals $I_{n}^{f^{m}}$ appear in quark loops during the renormalization of the Lagrangian and take the following form:
\begin{eqnarray}
	I_{n}^{f^{m}} =
	-i\frac{N_{c}}{(2\pi)^{4}}\int\frac{f^{m}(k^2_{\perp})}{(m^{2} - k^2)^{n}}\Theta(\Lambda_{3}^{2} - k^2_{\perp})
	\mathrm{d}^{4}k.
\end{eqnarray}
where $\Lambda_3=1030$~MeV is the three-dimensional cutoff parameter \cite{Volkov:2005kw}.

\section{The amplitude of the process $\tau \to \pi \pi \eta\nu_\tau$}

\begin{figure*}[t]
%\begin{figure*}[H]
 \centering
   \centering
   \begin{tikzpicture}
    \begin{feynman}
      \vertex (a) {\(\tau\)};
      \vertex [dot, below right=1.8cm of a] (b){};
      \vertex [below left=1.8cm of b] (c) {\(\nu_{\tau}\)};
      \vertex [blob, right=1.4cm of b] (d) {};
      \vertex [above right=1.8cm of d] (e) {\(\pi\)};
      \vertex [below right=1.8cm of d] (h) {\(\pi\)};
      \vertex [right=1.6cm of d] (f) {\(\eta (\eta')\)};
      \diagram* {
         (a) -- [fermion] (b),
         (b) -- [fermion] (c),
         (b) -- [boson, edge label'=\({W}\)] (d),    
         (d) -- [double] (e),  
         (d) -- [double] (h),
         (d) -- [double] (f),
      };
     \end{feynman}
    \end{tikzpicture}
   \caption{The contact diagram of the decays $\tau \to 2\pi \eta(\eta') \nu_\tau$. The shaded circle represents the sum of subdiagrams shown in Fig.~\ref{subdiagrams}.}
 \label{diagram1}
\end{figure*}
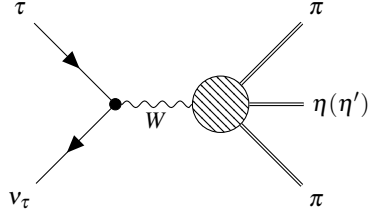

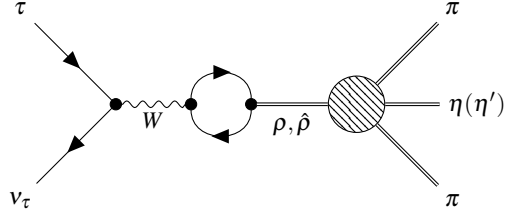
\begin{figure*}[t]
%\begin{figure*}[H]
 \centering
   \centering
   \begin{tikzpicture}
    \begin{feynman}
      \vertex (a) {\(\tau\)};
      \vertex [dot, below right=1.8cm of a] (b){};
      \vertex [below left=1.8cm of b] (c) {\(\nu_{\tau}\)};
      \vertex [dot, right=1.0cm of b] (d) {};
      \vertex [dot, right=0.8cm of d] (l) {};
      \vertex [blob, right=1.4cm of l] (g) {};
      \vertex [above right=1.8cm of g] (e) {\(\pi\)};
      \vertex [below right=1.8cm of g] (h) {\(\pi\)};
      \vertex [right=1.6cm of g] (f) {\(\eta (\eta')\)};
      \diagram* {
         (a) -- [fermion] (b),
         (b) -- [fermion] (c),
         (b) -- [boson, edge label'=\({W}\)] (d),
         (d) -- [fermion, inner sep=1pt, half left] (l),
         (l) -- [fermion, inner sep=1pt, half left] (d),
         (l) -- [double, edge label'=\({\rho, \hat{\rho}} \)] (g),
         (g) -- [double] (e),  
         (g) -- [double] (h),
         (g) -- [double] (f),
      };
     \end{feynman}
    \end{tikzpicture}
   \caption{The diagram with the intermediate vector mesons $\rho$ and $\hat{\rho}=\rho(1450)$ for $\tau$ decays. The shaded circle represents the sum of subdiagrams shown in Fig.~\ref{subdiagrams}.}
 \label{diagram2}
\end{figure*}

The diagrams of the process $\tau \to \pi \pi \eta\nu_\tau$ are given in Figs.~\ref{diagram1},~\ref{diagram2}.

The amplitude of this process in the NJL model takes the following form:
\begin{eqnarray}
	\mathcal{M} & = & 8 G_F V_{ud} m \left\{\mathcal{M}_{c\rho} + \mathcal{M}_{\rho\rho} + \mathcal{M}_{c\hat{\rho}} + \mathcal{M}_{\rho\hat{\rho}} + \mathcal{M}_{\hat{\rho}\rho} + \mathcal{M}_{\hat{\rho}\hat{\rho}} \right. \nonumber\\
    && \left. + \mathcal{M}_{c b} + \mathcal{M}_{\rho b} + \mathcal{M}_{\hat{\rho} b}\right\} L^\mu e_{\mu \nu \lambda \delta} p_{\eta}^\nu p_{\pi^-}^\lambda p_{\pi^0}^\delta,
\end{eqnarray}
where $G_F$ is the Fermi constant, $V_{ud}$ is the element of the Cabibbo–Kobayashi–Maskawa matrix, $L^\mu$ is the weak lepton current, $p_\eta$, $p_{\pi^-}$ and $p_{\pi^0}$ are the momenta of the final mesons.

\begin{figure*}[t]
%\begin{figure*}[H]
 \centering
  \begin{subfigure}{0.5\textwidth}
   \centering
   \begin{tikzpicture}
    \begin{feynman}
      \vertex [dot] (a) {};
      \vertex [dot, above right=1.8cm of a] (c){}; 
      \vertex [dot, below right=1.8cm of a] (e){};
      \vertex [dot, right=2.4cm of a] (d) {};
      \vertex [right=1.4cm of c] (f) {\(\pi\)};
      \vertex [right=1.4cm of d] (g) {\(\eta (\eta')\)};
      \vertex [right=1.4cm of e] (h) {\(\pi\)};
      \diagram* {
         (a) -- [fermion] (c),
         (c) -- [fermion] (d),
         (d) -- [fermion] (e),
         (e) -- [fermion] (a),
         (c) -- [double] (f),
         (d) -- [double] (g),
         (e) -- [double] (h),
      };
     \end{feynman}
    \end{tikzpicture}
  \end{subfigure}%
% \newline
 \centering
 \begin{subfigure}{0.5\textwidth}
  \centering
   \begin{tikzpicture}
     \begin{feynman}
      \vertex [dot] (d) {};      
      \vertex [dot, above right=1.4cm of d] (e) {};
      \vertex [dot, below right=1.4cm of d] (h) {};
      \vertex [dot, right=1.2cm of e] (f) {};
      \vertex [dot, above right=1.2cm of f] (n) {};  
      \vertex [dot, below right=1.2cm of f] (m) {};   
      \vertex [right=1.2cm of n] (l) {\(\ \pi \)}; 
      \vertex [right=1.2cm of m] (s) {\( \pi \)};  
      \vertex [right=1.4cm of h] (k) {\(\ \eta (\eta') \)}; 
      \diagram* {         
         (d) -- [fermion] (e),  
         (e) -- [fermion] (h),
         (d) -- [anti fermion] (h),
         (e) -- [double, edge label'=\({\rho, \hat{\rho}} \)] (f),
         (f) -- [fermion] (n),
         (n) -- [fermion] (m),
         (f) -- [anti fermion] (m), 
         (h) -- [double] (k),
         (n) -- [double] (l),
	 (m) -- [double] (s),
      };
     \end{feynman}
    \end{tikzpicture}
  % \caption{Д43}
  \end{subfigure}%
 \caption{The vertices $V\pi\pi\eta(\eta')$ with the box quark diagram and two triangle quark diagrams connected by the virtual vector mesons $\rho$ and $\hat{\rho}$.}
 \label{subdiagrams}
\end{figure*}
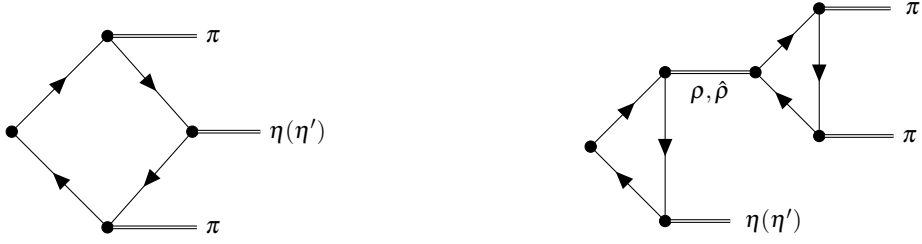%

The terms $\mathcal{M}_{\rho\rho}$, $\mathcal{M}_{\rho\hat{\rho}}$, $\mathcal{M}_{\hat{\rho}\rho}$ and $\mathcal{M}_{\hat{\rho}\hat{\rho}}$ describe the contributions from the diagrams with the intermediate $\rho$ mesons in the ground and first radially excited states. The first and second indices refer to the first and second intermediate states. The index $b$ denotes a box diagram. The index $c$ denotes the contact contribution. These terms take the following form:
\begin{eqnarray}
	\mathcal{M}_{c\rho} & = & 4 I_{3}^{\rho\eta} I_{2}^{\rho\pi\pi} Z_{\rho\pi\pi} BW_\rho^{p_{\pi^-} + p_{\pi^0}}, \nonumber\\
    \mathcal{M}_{\rho\rho} & = & 4\frac{C_\rho}{g_\rho} I_{3}^{\rho\rho\eta} I_{2}^{\rho\pi\pi} Z_{\rho\pi\pi} q^2 BW_\rho^q BW_\rho^{p_{\pi^-} + p_{\pi^0}}, \nonumber\\
    \mathcal{M}_{c\hat{\rho}} & = & 4 I_{3}^{\hat{\rho}\eta} I_{2}^{\hat{\rho}\pi\pi}Z_{\hat{\rho}\pi\pi} BW_{\hat{\rho}}^{p_{\pi^-} + p_{\pi^0}}, \nonumber\\
    \mathcal{M}_{\rho\hat{\rho}} & = & 4\frac{C_\rho}{g_\rho} I_{3}^{\rho\hat{\rho}\eta} I_{2}^{\hat{\rho}\pi\pi} Z_{\hat{\rho}\pi\pi} q^2 BW_\rho^q BW_{\hat{\rho}}^{p_{\pi^-} + p_{\pi^0}}, \nonumber\\
    \mathcal{M}_{\hat{\rho}\rho} & = & 4\frac{C_{\hat{\rho}}}{g_\rho} I_{3}^{\rho\hat{\rho}\eta} I_{2}^{\rho\pi\pi} Z_{\rho\pi\pi} q^2 BW_{\hat{\rho}}^q BW_\rho^{p_{\pi^-} + p_{\pi^0}}, \nonumber\\
    \mathcal{M}_{\hat{\rho}\hat{\rho}} & = & 4\frac{C_{\hat{\rho}}}{g_\rho} I_{3}^{\hat{\rho}\hat{\rho}\eta} I_{2}^{\hat{\rho}\pi\pi} Z_{\hat{\rho}\pi\pi} q^2 BW_{\hat{\rho}}^q BW_{\hat{\rho}}^{p_{\pi^-} + p_{\pi^0}}, \nonumber\\
    \mathcal{M}_{c b} & = & I_4^{\pi\pi\eta}, \nonumber\\
    \mathcal{M}_{\rho b} & = & \frac{C_\rho}{g_\rho} I_4^{\rho\pi\pi\eta} q^2 BW_\rho^q, \nonumber\\
    \mathcal{M}_{\hat{\rho} b} & = & \frac{C_{\hat{\rho}}}{g_\rho} I_4^{\hat{\rho}\pi\pi\eta} q^2 BW_{\hat{\rho}}^q,
\end{eqnarray}
where
\begin{eqnarray}
	Z_{\rho\pi\pi} = 1 - 4\frac{I_{2}^{\rho a_1 \pi}I_{2}^{a_1\pi}}{I_{2}^{\rho\pi\pi}} \frac{m^2}{M_{a_1}^2}.
\end{eqnarray}

%Здесь промежуточные состояния описываются пропагаторами Брейта-Вигнера:
Here the intermediate states are described by the Breit-Wigner propagator
\begin{eqnarray}
	BW_{meson}^p = \frac{1}{M_{meson}^2 - p^2 - i\sqrt{p^2}\Gamma_{meson}}.
\end{eqnarray}

%Интегралы, возникающие в амплитуде, принимают следующий вид:
The integrals that arise in the amplitude take the form
\begin{eqnarray}
	I_{n}^{meson_1, meson_2, \dots, \hat{meson_1}, \hat{meson_2}, \dots} =
	-i\frac{N_{c}}{(2\pi)^{4}}\int\frac{A_{meson_1}A_{meson_2}\dots B_{meson_1}B_{meson_2}\dots}{(m^{2} - k^2)^{n}}\Theta(\Lambda_{3}^{2} - k^2_{\perp})
	\mathrm{d}^{4}k,
\end{eqnarray}
where $A$ and $B$ are the factors from the Lagrangian (\ref{Lagrangian}) relating the quark and meson fields and containing the mixing angles.
%где $A$ и $B$ --- множители из Лагранжиана (\ref{Lagrangian}), связывающие кварковые и мезонные поля и содержащие углы смешивания.

The branching fraction of the process calculated using the amplitude given above turns out to be lower than the experimental data presented in the PDG:
\begin{eqnarray}
	Br(\tau \to \pi \pi \eta\nu_\tau)_{NJL} & = & (0.75 \pm 0.11) \times 10^{-3}, \\
    Br(\tau \to \pi \pi \eta\nu_\tau)_{exp} & = & (1.39 \pm 0.07) \times 10^{-3} \textrm{ \cite{ParticleDataGroup:2024cfk}}
\end{eqnarray}

The theoretical uncertainty of the NJL model can be estimated approximately at the level of 15\%\cite{Volkov:2017arr,Volkov:2022jfr}.

According to the works \cite{Achasov:2017kqm,BaBar:2018erh}, the phase factors $e^{i\phi}$ with $\phi = 180^{\circ}$ describing the phase shift between the ground and first radially excited states are of great importance. However, this factor is not described by the NJL model and can be included on the basis of experimental data by redefining the Breit-Wigner propagator for excited meson states:
\begin{eqnarray}
	BW_{\rho'}^p \to e^{i\phi} BW_{\rho'}^p.
\end{eqnarray}
Then the absence of this factor is appropriate for the case $\phi = 0^{\circ}$.

While taking into account $e^{i\pi}$, the branching fraction of this process in the NJL model turns out to be higher than the average value given in the PDG but close to recent BaBar and CMD-3 data
\begin{eqnarray}
    Br(\tau \to \pi \pi \eta\nu_\tau)_{NJL} & = & (1.84 \pm 0.27) \times 10^{-3} \quad(\phi = 180^{\circ}). \nonumber\\
    Br(\tau \to \pi \pi \eta\nu_\tau)_{BaBar} & = & (1.63 \pm 0.08) \times 10^{-3} \textrm{ \cite{BaBar:2018erh}}, \nonumber\\
    Br(\tau \to \pi \pi \eta\nu_\tau)_{CMD-3} & = & (1.68 \pm 0.17) \times 10^{-3} \textrm{ \cite{Gribanov:2019qgw}}.
\end{eqnarray}

%It is interesting to note that in the later experiments the following results have been obtained:
%\begin{eqnarray}
 %   Br(\tau \to \pi \pi \eta\nu_\tau)_{exp} & = & (1.68 \pm 0.17) \times 10^{-3} \textrm{ \cite{Gribanov:2019qgw}}, \nonumber\\
  %  Br(\tau \to \pi \pi \eta\nu_\tau)_{exp} & = & (1.63 \pm 0.08) \times 10^{-3} \textrm{ \cite{BaBar:2018erh}}.
%\end{eqnarray}
The theoretical result obtained with $e^{i\pi}$ is in agreement with the experimental data \cite{Gribanov:2019qgw,BaBar:2018erh} in the framework of the theoretical and experimental uncertainties. Note that the inclusion of the phase factor previously led to a noticeable improvement in the results obtained for the processes of $\tau$-decays and $e^+e^-$ annihilation containing intermediate channels with radially excited mesons \cite{Volkov:2017arr}.

The structure of the amplitude of the decay $\tau \to \pi \pi \eta'\nu_\tau$ almost coincides with the structure of the amplitude of the decay $\tau \to \pi \pi \eta\nu_\tau$ with the replacement of vertices related to the $\eta$ meson by the vertices related to the $\eta'$ meson under the integrals. 

The experimental limit on the branching fraction of this process is~\cite{ParticleDataGroup:2024cfk}
\begin{eqnarray}
	Br(\tau \to \pi \pi \eta'\nu_\tau)_{PDG} & < & 1.2 \times 10^{-5}. 
\end{eqnarray}

The theoretical results in the case $\phi = 0^{\circ}$ satisfy this constraint
\begin{eqnarray}
	Br(\tau \to \pi \pi \eta'\nu_\tau) & = & (0.32 \pm 0.04) \times 10^{-5} \quad (\phi = 0^{\circ}). 
\end{eqnarray}

However, the use of the phase $\phi = 180^{\circ}$ leads to exceeding the experimental threshold
\begin{eqnarray}
	Br(\tau \to \pi \pi \eta'\nu_\tau) & = & (2.45 \pm 0.36) \times 10^{-5} \quad (\phi = 180^{\circ} ).
\end{eqnarray}

\section{The decay $\tau \to \pi^-2\eta \nu_{\tau}$}
\label{pi2eta}
%Амплитуда распада $\tau \to \pi^-2\eta\nu_{\tau}$ состоит из вкладов промежуточных аксиально-векторного и псевдоскалярного каналов в основном и первом радиально-возбуждённом состоянии. Для этих каналов мы учитываем в качестве второго промежуточного состояния $a_0(980)$. Полная амплитуда распада может быть записана в виде
The total decay amplitude of $\tau \to \pi^-2\eta\nu_{\tau}$ consists of the contributions from the intermediate axial-vector and pseudoscalar channels in the ground and first radially excited states. For these channels, we take into account $a_0(980)$ as the second intermediate state. 
The total decay amplitude can be written as
\begin{eqnarray}
\label{amplitude4}
\mathcal{M}(\tau \to \pi^- 2\eta\nu_\tau) & = &64 G_{F} V_{ud} L_{\mu} \left\{ \mathcal{M}_{ca_0} + \mathcal{M}_{a_1a_0}+ \mathcal{M}_{\hat{a_1}a_0} + \mathcal{M}_{\pi a_0} + \mathcal{M}_{\hat{\pi} a_0} \right\}^{\mu},
\end{eqnarray}
%где адронная часть амплитуды представлена в фигурных скобках. Здесь первые три слагаемые соответствуют вкладам от аксиально-векторных каналов, т.е. контактному каналу и каналу с аксиально-векторнымы мезонами $a_1$ и $\hat{a_1}$ в первом промежуточном состоянии. 
where the hadronic part of the amplitude is presented in curly brackets. Here the first three terms correspond to the contributions from the axial-vector channels, i.e. the contact channel and the channel with the axial-vector mesons $a_1$ and $\hat{a_1}$ in the first intermediate state.
The last two terms correspond to the pseudoscalar channels. In all channels, the scalar meson $a_0(980)$ is represented as the second intermediate state, which decays into final products through the strong vertex $a_0 \pi \eta$. The expressions for the hadronic parts of this amplitude of the contact, axial-vector, and pseudoscalar channels have the form 
%Последние два слагаемых соответствуют псевдоскалярным каналам. Во всех каналах в качестве второго промежуточного состояния выступает скалярный мезон $a_0(980)$, который через сильную вершину $a_0 \pi \eta$ распадается на конечные продукты.    Выражения для адронных частей данной амплитуды контактных, аксиально-векторных и псевдосклярных каналов имеют вид
\begin{eqnarray}
    \mathcal{M}_{ca_0} & = & m_u I_{2}^{a_0\pi\eta} I_{2}^{a_0\eta} Z_{a_1\pi\eta} BW_{a_0}^{p_{\pi^-} + p^{(1)}_{\eta}} g_{\mu \nu} (p_{\pi^-}+p_{\eta^{(1)}}-p_{\eta^{(2)}})_\nu + (p_{\eta^{(1)}} \leftrightarrow p_{\eta^{(2)}}), \nonumber\\
    \mathcal{M}_{a_1a_0} & = & m_u \frac{C_{a_1}}{g_\rho} I_{2}^{a_0\pi\eta} I_{2}^{a_1a_0\eta} Z_{a_1\pi\eta} BW_{a_0}^{p_{\pi^-} + p^{(1)}_{\eta}} \left( g_{\mu \nu} (q^2 - 6 m^2_u) - q_\mu q_\nu \right) \times
     \nonumber\\
    &&
    BW_{a_1}^{q}(p_{\pi^-}+p_{\eta^{(1)}}-p_{\eta^{(2)}})_\nu 
    + (p_{\eta^{(1)}} \leftrightarrow p_{\eta^{(2)}}), \nonumber\\
    \mathcal{M}_{\hat{a_1}a_0} & = & m_u \frac{C_{\hat{a_1}}}{g_\rho} I_{2}^{a_0\pi\eta} I_{2}^{\hat{a_1}a_0\eta} Z_{a_1\pi\eta} BW_{a_0}^{p_{\pi^-} + p^{(1)}_{\eta}} \left( g_{\mu \nu} (q^2 - 6 m^2_u) - q_\mu q_\nu \right) \times
     \nonumber\\
    &&
    BW_{\hat{a_1}}^{q}(p_{\pi^-}+p_{\eta^{(1)}}-p_{\eta^{(2)}})_\nu + (p_{\eta^{(1)}} \leftrightarrow p_{\eta^{(2)}}), \nonumber\\    
    \mathcal{M}_{\pi a_0} & = & 4 F_{\pi} I_{2}^{a_0\pi\eta} I_{2}^{a_0\pi\eta}Z_{\pi\pi\eta} BW_{\pi}^{q} BW_{a_0}^{p_{\pi^-} + p^{(1)}_{\eta}} q^\mu + (p_{\eta^{(1)}} \leftrightarrow p_{\eta^{(2)}}), \nonumber\\
    \mathcal{M}_{\hat{\pi} a_0} & = & 4 F_{\pi} I_{2}^{\hat{\pi}a_0\eta} I_{2}^{a_0\pi\eta}Z_{\hat{\pi}\pi\eta} BW_{\hat{\pi}}^{q} BW_{a_0}^{p_{\pi^-} + p^{(1)}_{\eta}} q^\mu + (p_{\eta^{(1)}} \leftrightarrow p_{\eta^{(2)}}),
\end{eqnarray}
where $q=p_{\pi}+p_{\eta^{(1)}}+p_{\eta^{(2)}}$. Here the factors $Z_{a_1\pi\eta}$ and $Z_{\pi\pi\eta}$ arise when taking into account the non-diagonal $\pi-a_1$ and $\eta-f_1$ transitions. The expressions for these factors take the form
\begin{eqnarray}
    Z_{a_1\pi\eta} = 1 - \frac{I_{2}^{a_1 a_0 \eta}I_{2}^{a_1\pi}}{I_{2}^{a_0\pi\eta}} \frac{{\left(p_{\pi^-} + p^{(1)}_{\eta} \right)}^2 - M^2_\eta}{M_{a_1}^2} - \frac{I_{2}^{f_1 a_0 \pi}I_{2}^{f_1\eta}}{I_{2}^{a_0\pi\eta}} \frac{{\left(p_{\pi^-} + p^{(1)}_{\eta} \right)}^2 - M^2_\eta}{M_{f_1}^2},
\end{eqnarray}
\begin{eqnarray}
    Z_{\pi\pi\eta} & = & \left( 1- \frac{I_{2}^{f_1 a_0 \pi}I_{2}^{f_1\eta}}{I_{2}^{a_0\pi\eta}} \frac{{q^2 - \left(p_{\pi^-} + p^{(1)}_{\eta} \right)}^2 }{M_{f_1}^2} - \frac{I_{2}^{a_1 a_0 \eta}I_{2}^{a_1\pi}}{I_{2}^{a_0\pi\eta}} \frac{{\left(p_{\pi^-} + p^{(1)}_{\eta} \right)}^2 - M^2_\eta }{M_{a_1}^2}
    \right) \times
 \nonumber\\
    &&
    \left( 1- \frac{I_{2}^{a_1 a_0 \eta}I_{2}^{f_1\eta}}{I_{2}^{a_0\pi\eta}} \frac{{\left(p_{\pi^-} + p^{(1)}_{\eta} \right)}^2 - M^2_\eta }{M_{f_1}^2} - \frac{I_{2}^{f_1 a_0 \pi}I_{2}^{f_1\eta}}{I_{2}^{a_0\pi\eta}} \frac{{\left(p_{\pi^-} + p^{(1)}_{\eta} \right)}^2 - M^2_\pi }{M_{f_1}^2}
    \right).
\end{eqnarray}

%В результате, для парциальной ширины распада $\tau \to \pi^-2\eta \nu_{\tau}$ получаем следующие оценки в расширенной НИЛ модели 
As a result, in the extended NJL model for the branching fractions of decay $\tau \to \pi^-2\eta \nu_{\tau}$ we obtain
\begin{eqnarray}
    Br(\tau \to \pi \eta \eta\nu_\tau) & = & (6.0 \pm 0.9) \times 10^{-6}, \quad (\phi = 180^{\circ}) \nonumber\\
    Br(\tau \to \pi \eta \eta\nu_\tau) & = & (8.80 \pm 1.32) \times 10^{-6}, \quad (\phi = 0^{\circ}).
\end{eqnarray}

The experiment only provides an upper limit on the decay branching fractions \cite{ParticleDataGroup:2024cfk}
%Эксперимент дает только ограничение на верхний предел ширины распада \cite{}   
\begin{eqnarray}
    Br(\tau \to \pi \eta \eta\nu_\tau)_{PDG} & = & < 7.4 \times 10^{-6}.
\end{eqnarray}

The result in the NJL model for the phase $\phi = 180^{\circ}$ satisfies these data.
%Результат в НИЛ модели для фазы $\phi = 180^{\circ}$ удовлетворяет этому ограничению.

The decay amplitude of $\tau \to \pi \eta \eta' \nu_\tau$ can be obtained from formula (\ref{amplitude4}) with the replacement of the vertices $\eta \to \eta'$ and the mass $M_\eta \to M_{\eta'}$, respectively. The calculations within the extended NJL model lead to the following result for the decay branching fractions $\tau \to \pi \eta \eta' \nu_\tau$

%Амплитуду распада $\tau \to \pi \eta \eta' \nu_\tau$ можно получить в форме (\ref{amplitude4}) с заменой вершин $\eta \to \eta'$ и массы $M_\eta \to M_{\eta'}$, соответственно. Выполненные расчеты в расширенной модели НИЛ приводят к следующему результату для ширины распада $\tau \to \pi \eta \eta' \nu_\tau$ 
\begin{eqnarray}
    Br(\tau \to \pi \eta \eta'\nu_\tau) & = & (4.0 \pm 0.6) \times 10^{-7}.
\end{eqnarray}

\section{Conclusion}
In this paper, three-meson $\tau$ decays with the participation of $\eta/\eta'$ mesons in the final state have been considered within the extended NJL model. The amplitudes for the contact channels and channels with intermediate vector, axial-vector and pseudoscalar mesons in the ground and first radially excited states were calculated. The results are obtained with the phase angle between the ground and first radially excited meson states. In the case of the $\tau \to \pi \pi \eta\nu_\tau$ process, the angle value $\phi = 180^{\circ}$ leads to a result that does not correspond to the PDG data but is consistent with the recent results of the BaBar and CMD-3 collaborations \cite{Gribanov:2019qgw,BaBar:2018erh}.
Calculation of the process $\tau \to \pi \pi \eta'\nu_\tau$ with a given value of the angle leads to exceeding the threshold given in the PDG. However, there is reason to believe that further experimental study of this process may lead to a correction of this threshold, since analysis of the latest BaBar and CMD-3 data provides branching fractions somewhat larger than the average value given in PDG.

The decays $\tau \to \pi \pi \eta\nu_\tau$ and $\tau \to \pi \pi \eta'\nu_\tau$ occur only through vector channels. These decays were previously described in the NJL model in the paper~\cite{Volkov:2013zba}. However, the decay width can be calculated based on the $e^+ e^-$ annihilation process using the phase volume transformation. Similar calculations of the decays $\tau \to \pi \pi \eta(\eta')\nu_\tau$ within the vector current conservation hypothesis (CVC) were performed in the paper \cite{Cherepanov:2011zz} where the decay widths were estimated as $Br(\tau \to \pi \pi \eta\nu_\tau) = (1.53 \pm 0.18) \times 10^{-3}$ and $Br(\tau \to \pi \pi \eta'\nu_\tau) < 3.2 \times 10^{-5}$.

The $\tau \to \pi \pi \eta\nu_\tau$ process was considered in the theoretical work~\cite{GomezDumm:2012dpx} within the chiral perturbation theory with resonances. However, the calculations were performed without taking into account the contribution of the intermediate radially excited $\rho(1450)$ meson. The spectral function for the invariant mass $M_{\pi\pi\eta}$ was fitted by model parameters based on Belle data \cite{Belle:2008jjb}.
At the same time, the calculations in the present work in the NJL model show that the contribution from the intermediate $\rho(1450)$ meson is important for describing the decay widths.

%В работе ~\cite{GomezDumm:2012dpx} также был рассмотрен распад $\tau \to \pi \pi \eta'\nu_\tau$. Однако теоретическая оценка для парциальной ширины данного распада превышает экспериментальный порог. Важно отметить, что оценки для распада $\tau \to \pi \pi \eta'\nu_\tau$, полученные в НИЛ модели и с помощью гипотезы CVC также превышают экспериментальный предел для ширины распада. Всё это может быть указанием на необходимость более тщательного экспериментального и теоретического исследования данного процесса. 
In the paper ~\cite{GomezDumm:2012dpx} the $\tau \to \pi \pi \eta'\nu_\tau$ decay was also considered. However, the theoretical estimate for the partial width of this decay is in the range $Br(\tau \to \pi \pi \eta'\nu_\tau)=(1-4.5) \times 10^{-4}$ and exceeds the experimental threshold. It is important to note that the estimates for the $\tau \to \pi \pi \eta'\nu_\tau$ decay obtained using the CVC hypothesis and within the NJL model exceed the experimental limit for the decay width. All this may indicate the need for a more thorough experimental and theoretical study of this process.

The results of calculations in the NJL model for the decays $\tau \to \pi \eta\eta(\eta')\nu_\tau$ are given in the chapter \ref{pi2eta}. In contrast to the above processes, the matrix elements of these decays are determined by the axial current.  The calculations show that the main contribution to the decay width of $\tau \to \pi \eta \eta\nu_\tau$ is given by the axial-vector channels with the intermediate states $a_1(1260)$ and $a_1(1640)$ containing the scalar meson $a_0(980)$. Note that among the numerous $\tau$ decays, the decay $\tau \to \pi \eta\eta\nu_\tau$ is unique where the scalar isovector meson $a_0$ plays a dominant role in determining the decay width.

In a recent paper \cite{Volkov:2024wsu}, a theoretical estimation of the contribution of the intermediate scalar meson $f_0(500)$ to the total decay width $\tau \to \pi^-\pi^0\pi^0\nu_\tau$ was given. It was shown that the contribution of the intermediate channel $f_0(500)\pi$ is 13\% of the total decay width, which is consistent with the CLOE II data $(16.18 \pm 3.85)$ \% \cite{CLEO:1999rzk}. Moreover, taking into account the scalar channel in this decay turns out to be important for a correct description of the invariant mass distribution function of three pions . All this demonstrates the importance of taking into account channels with intermediate scalar mesons for a number of three-meson $\tau$ decays.

\subsection*{Acknowledgments}
The authors thank Professor A. B. Arbuzov for useful discussions.

\end{document}